\documentclass[preprint,12pt]{elsarticle}

\usepackage{algorithm2e}
\usepackage{algpseudocode}
\usepackage{amssymb}
\usepackage{xcolor}

\begin{document}

\begin{frontmatter}



\title{A High-Performance Implementation of Atomistic Spin Dynamics Simulations on x86 CPUs}




\author[neu,sf,slac]{Hongwei Chen}
\author[ucr]{Yujia Zhai}
\author[sf,slac]{Joshua J. Turner}
\author[neu]{Adrian Feiguin}

\affiliation[neu]{organization={Department of Physics, Northeastern University},
            city={Boston},
            postcode={02115}, 
            state={MA},
            country={USA}}
\affiliation[ucr]{organization={Department of Computer Science and Engineering, University of California, Riverside},
            city={Riverside},
            postcode={92501}, 
            state={CA},
            country={USA}}

\affiliation[sf]{organization={Stanford Institute for Materials and Energy Sciences, Stanford University},
            city={Stanford},
            postcode={94305}, 
            state={CA},
            country={USA}}
\affiliation[slac]{organization={Linac Coherent Light Source, SLAC National Accelerator Laboratory},
            city={Menlo Park},
            postcode={94720}, 
            state={CA},
            country={USA}}

\begin{abstract}
Atomistic spin dynamics simulations provide valuable information about the energy spectrum of magnetic materials in different phases, allowing one to identify instabilities and the nature of their excitations. However, the time cost of evaluating the dynamical correlation function $S(\mathbf{q}, t)$ increases quadratically as the number of spins $N$,  leading to significant computational effort, making the simulation of large spin systems very challenging. In this work, we propose to use a highly optimized general matrix multiply (GEMM) subroutine to calculate the dynamical spin-spin correlation function that can achieve near-optimal hardware utilization. Furthermore, we fuse the element-wise operations in the calculation of $S(\mathbf{q}, t)$ into the in-house GEMM kernel, which results in further performance improvements of 44\% - 71\% on several relatively large lattice sizes when compared to the implementation that uses the GEMM subroutine in OpenBLAS, which is the state-of-the-art open source library for Basic Linear Algebra Subroutine (BLAS). 

\end{abstract}




\end{frontmatter}

\section{Introduction}

The control and manipulation of the electronic spin degree of freedom in magnetic materials offer the promise of novel technologies that hold many advantages, such as low energy consumption, low heat dissipation, and persistent memories. Recent developments in the burgeoning field of ``spintronics'' \cite{Bader2010, Duine2011, Sinova2012, Baltz2018, HIROHATA2020, Avsar2020} have focused on improving the efficiency and reliability of spin-based devices, as well as exploring new materials and techniques for manipulating spins. These advances have led to the development of spin-based transistors, magnetic memories, and sensors. Unlike the charge degree of freedom, the spin behaves as a vector allowing one to explore a higher dimensional realm of possible magnetic states with exotic excitations. This is illustrated by the recent discovery of novel types of magnetic textures dubbed ``skyrmions''\cite{Marrows2021} which, unlike domain wall defects (``solitons''), are like vortices with topological properties that make them very robust at room temperature and can be used to store and transport information. 

The study of this field has been guided by theory and numerical simulations using Monte Carlo (MC) and molecular dynamics-like (MD) methods \cite{takahashi1990dynamics, chen1994spin, samarakoon2017comprehensive, zhang2019dynamical, mohanta2020signatures, saha2021spin} to understand time-dependent non-equilibrium properties of magnetic phases and defects. These techniques are able to simulate the behavior of thousands of spins with unprecedented accuracy and faithfully reproduce experimental observations. An important measurement that provides information about the spectrum and the nature of the excitations is provided by the spin correlations and, in particular, the dynamic structure factor $S(\mathbf{q}, \omega)$ that can be readily compared to experimental probes such as inelastic neutron and inelastic x-ray scattering \cite{Ament-2011-RMP}, or their time-domain counterpart $S(\mathbf{q}, t)$, through methods such as neutron spin echo or x-ray photon correlation spectroscopy \cite{MarkSutton-2008,sinha-advmat-2014}.    

Some previous research works studied the parallel implementation of the Monte Carlo method \cite{kaupuvzs2010parallelization, weigel2011gpu, komura2016improved, liang2017gpu, hassani2018parallelization} or molecular dynamics solutions of the Landau-Lifshitz-Gilbert (LLG) equations\cite{evans2014atomistic, etz2015atomistic, muller2019spirit} on multi-processor hardware. These techniques have been included in some open-source packages for spin dynamics simulation such as UppASD\cite{skubic2008method}, Vampire\cite{evans2014atomistic}, Spirit\cite{muller2019spirit}, and Sunny\cite{sunny}. However, the computational cost for obtaining the dynamic correlation function $S(\mathbf{q}, t)$ becomes the main bottleneck (see discussion below), significantly limiting the system sizes one can simulate.  In this paper, we explore the high-performance implementation of atomistic spin dynamics simulations on x86 CPUs, finding that the GEMM subroutine can greatly accelerate the calculation of $S(\mathbf{q}, t)$. Furthermore, we improve the performance by fusing the element-wise operation into our in-house GEMM subroutine to hide the memory latency. Our in-house GEMM subroutine has a similar performance compared to OpenBLAS\cite{xianyi2012model, wang2013augem} on AVX2-enabled processors. In our test, the kernel fusion strategy combined with highly optimized GEMM can speed up the $S(\mathbf{q}, t)$ calculations up to 8-fold on the Intel platform. 

\section{Numerical Method}
Magnetic materials can be modeled by the following generalized Heisenberg Hamiltonian
\begin{equation} \label{eq:hamiltonian}
\mathcal{H} = \sum_{<ij>}J_{ij} \mathbf{S}_i\cdot \mathbf{S}_j -\sum_{<ij>}\mathbf{D}_{ij}\cdot (\mathbf{S}_i \times \mathbf{S}_j)  - A \sum_i (S^z_i) ^2 - H_z\sum_i S^z_i
\end{equation} 
where $S_i$ is a spin vector in the unit sphere, $\mathbf{D}_{ij}$ is the Dzyaloshinskii–Moriya interaction vector, $A$ is the perpendicular anisotropy, and $H_z$ the perpendicular magnetic field. In this work, we consider an antiferromagnetic Heisenberg spin model on the 2D square lattice with only nearest-neighbor interaction, 
\begin{equation}
    \mathcal{H} = J \sum_{<ij>} \mathbf{S}_i\cdot \mathbf{S}_j, 
\end{equation}
where the coupling $J$ can be positive (antiferromagnetic), favoring anti-parallel arrangements of spins, or negative (ferromagnetic), favoring configuration in which the spins point in the same direction. For this calculation, we take $|J|=1$ without loss of generality. The finite temperature physics of this problem can be studied by means of classical Monte Carlo (MC) simulations using the Metropolis algorithm, and it has been widely used to find the ground state of the classical spin Hamiltonian for decades\cite{binder1973monte}. 

The procedure to perform classical Monte Carlo is well described in the literature\cite{binder1993monte, murthy2001introduction, landau_binder_2014}, and we hereby summarize it. Given a spin configuration, each spin is updated by proposing a new trial move  randomly chosen on the sphere; the new spin orientation will be accepted or rejected using the von Neumann rejection method according to the probability $\mathbb{P}(\Delta E)=e^{(-\beta \Delta E)}$, where $\Delta E$ is the energy difference between the new and old spin configurations, and $\beta=1/T$ is the inverse temperature. This procedure is guaranteed to obey detailed balance and samples an equilibrium finite temperature distribution consistent with the partition function of the model\cite{binder1993monte}. At low temperatures, the acceptance rate of the random walk may be very small. One common technique to overcome this issue is that the new trial spin was chosen randomly within a small cone spanned around the initial spin. In our implementation, we adaptively change the size of the cone to make sure that the acceptance rate is at least 20\%.  
 
To investigate the time-dependent dynamical behavior of classical spin models, Monte Carlo simulations and molecular dynamics techniques need to be combined. MC simulations are used to generate spin configurations corresponding to the equilibrium distribution at a specified temperature. The temporal evolution of classical spins is described by the Landau-Lifshitz-Gilbert equation 
\begin{equation} \label{eq:ll}
    \frac{ d \mathbf{S_i}(t)}{dt} = \frac{\partial \mathcal{H}}{\partial \mathbf{S_i}(t)} \times \mathbf{S_i}(t),
\end{equation}
where $\mathcal{H}$ is the Hamiltonian of the magnetic material and $\mathbf{S_i}(t)$ are the spin vectors. With the spin trajectories $\mathbf{S_i}(t)$, the dynamic correlation function can be computed as 
\begin{equation} \label{eq:sqt}
    S(\mathbf{q}, t) = \frac{1}{N}\sum_{\alpha} \sum_{\mathbf{r}, \mathbf{r'}}e^{-i\mathbf{q}\cdot (\mathbf{r} -\mathbf{r'})}  C^{\alpha}(\mathbf{r} -\mathbf{r'}, t), 
\end{equation}
which is the Fourier transformation of the 
time-dependent spin-spin correlation function $C^{\alpha}(\mathbf{r} - \mathbf{r'}, t)$  defined as 
\begin{equation} \label{eq:ssc}
    C^{\alpha}(\mathbf{r} - \mathbf{r'}, t) = [\langle S^\alpha_{\mathbf{r}}(t) S^\alpha_{\mathbf{r'}}(0) \rangle - \langle S^\alpha_{\mathbf{r}}(t)\rangle \langle S^\alpha_{\mathbf{r'}}(0)\rangle ]. 
\end{equation}
The $\langle \rangle$ symbol means the statistical average over thermalized realizations of the Monte Carlo simulation, and $\alpha$ represents the $x$, $y$, and $z$ components. The dynamic spin structure factor $S(\mathbf{q}, \omega)$ can then be obtained by Fourier transforming $S(\mathbf{q}, t)$ to the frequency domain.

In some previous studies, the Fourier transform (FT) was computed using the conventional definition in a finite time window, resulting in ringing oscillations along the $\omega$ axis in $S(\mathbf{q},\omega)$. These may be alleviated by a Hanning window, leading to a lineshape broadening inversely proportional to the width of the time window. However, in this work, we utilize the fast Fourier transform (FFT) to convert $S(\mathbf{q}, t)$ to the frequency domain. The FFT method was chosen because it is specifically designed to handle discrete signals, which have a finite number of samples, whereas the FT is designed for continuous signals. Another advantage of the FFT is that it allows for efficient computation of the frequency as it reduces the computational complexity from $O(n^2)$ to $O(n\log{n})$, where $n$ is the length of the discrete signal.

\section{Optimizations and Benchmarks}

In this section, we present a step-by-step description of the multiple optimization procedures we implement to speed up spin dynamics simulations using delicately tuned assembly kernels, and we compare them to an ordinary baseline version. To validate the effectiveness of our optimizations, we evaluate the performance on a 64-core AMD EPYC 7763 CPU, whose boost frequency is up to 3.5 GHz. The CPU node is connected to DDR4 memory systems at 3200 MHz. We compile programs using $\mathtt{gcc\ 11.2.0}$ with default compiler optimization flags turned on.

\subsection{Baseline}

\RestyleAlgo{ruled}

\begin{algorithm}[hbt!]
\caption{The baseline implementation for $S(\mathbf{q}, t)$}\label{alg:baseline}
\KwData {$\mathbf{S}_{i,r}(0)$ $M$ spin configurations from Monte Carlo ($1 \leq  i \leq M$) }
\For{$r' \gets 1$ \KwTo $N$} 
{
     $\langle S^\alpha_{r'}(0)\rangle \gets \mathbf{S}_{i,r'}(0)$ ($1 \leq  i \leq M$) \Comment{Calculate statistical average}
}
Rearrange $\mathbf{S}_{i,r}(0)$ \Comment{To access data continuously}

\For{$t \gets 1$ \KwTo $T$} 
{

\For{$i \gets 1$ \KwTo $M$}
{
    \For{$r \gets 1$ \KwTo $N$}
    {
         $\mathbf{S}_{i,r}(t) \gets \mathbf{S}_{i,r}(t-1) $ \Comment{Solve LLG equation}
    }
}

\For{$r \gets 1$ \KwTo $N$}
{
     $\langle S^\alpha_{r}(t)\rangle \gets \mathbf{S}_{i, r}(t)$ ($1 \leq  i \leq M$)  \Comment{Calculate statistical average}
}

\For{$r \gets 1$ \KwTo $N$}
{
     Load $\mathbf{S}_{i, r}(t)$ ($1 \leq  i \leq M$) to a new array
    
    \For{$r' \gets 1$ \KwTo $N$}
    {
         $C^{\alpha}(\mathbf{r} - \mathbf{r'}, t) \gets 0$
        
        \For{$k \gets 1$ \KwTo $M$}
        {
             $C^{\alpha}(\mathbf{r} - \mathbf{r'}, t) \  += \  S^\alpha_{\mathbf{r}}(t) S^\alpha_{\mathbf{r'}}(0) $ \Comment{Calculate $\langle S^\alpha_{\mathbf{r}}(t) S^\alpha_{\mathbf{r'}}(0) \rangle$}
        }
         $C^{\alpha}(\mathbf{r} - \mathbf{r'}, t) \ -= \langle S^\alpha_{\mathbf{r}}(t)\rangle \langle S^\alpha_{\mathbf{r'}}(0)\rangle $
         $S(\mathbf{q}, t) \gets e^{-i\mathbf{q}\cdot (\mathbf{r} -\mathbf{r'})}  C^{\alpha}(\mathbf{r} -\mathbf{r'}, t)$
    }        
}

}
\end{algorithm}

Prior to discussing the specific implementation, we first analyze the time complexity of the equations involved in the calculation. It can be easily seen that while solving the LLG equations at time $t$, the number of interactions for each site that must be accounted for is a constant for a given Hamiltonian, regardless of the system size. Hence, the time complexity for a single thermalized realization is $O(N)$, where $N$ denotes the system size. Calculating the dynamical correlation function $S(\mathbf{q}, t)$ is computationally more expensive due to the summation over $r$ and $r'$ in the dynamical spin-spin correlation function $C^{\alpha}(\mathbf{r} - \mathbf{r'}, t)$. Consequently, evaluating $S(\mathbf{q}, t)$ for $M$ Monte Carlo realizations leads to a time complexity of $O(M\cdot N^2)$, while solving the LLG equations has a time complexity of $O(M\cdot N)$.

To parallelize the implementation of solving the LLG equations on a multi-core CPU, a similar approach to parallel Monte Carlo is used.  The time evolution in Eq.(\ref{eq:ll}) was computed using the fourth-order Runge-Kutta method.  While solving $\frac{ d \mathbf{S_i}(t)}{dt}$, the value of $\frac{ d \mathbf{S_i}(t)}{dt}$ is stored in a separate array to enable its simultaneous evaluation on different processors. Thus, in our implementation, we distribute the workload evenly among $N_p$ processors using $\mathtt{OpenMP}$. For $M$ Monte Carlo realizations, we can either solve the LLG equation sequentially for each realization or concatenate the spin configurations and solve the equation once. Our tests show that the latter method is significantly faster since it avoids the overhead introduced by the system calls in $\mathtt{OpenMP}$. Therefore, we adopt the second method for our following simulations and discussions.

As for the calculation of $S(\mathbf{q},t)$, three nested $\mathtt{for}$ loops are required. The outer two loops are used to iterate over the indices $r$ and $r'$, while the inner loop is necessary to compute the statistical average for the term $\langle S^\alpha_{\mathbf{r}}(t) S^\alpha_{\mathbf{r'}}(0) \rangle$. The computation of $C^{\alpha}(\mathbf{r} - \mathbf{r'}, t)$ and $e^{-i\mathbf{q}\cdot (\mathbf{r} -\mathbf{r'})} C^{\alpha}(\mathbf{r} -\mathbf{r'}, t)$ can then be carried out in the second loop. The implementation described above represents a straightforward approach to evaluate $S(q,t)$, but optimizing its performance is essential for the simulation due to the high computational cost. Firstly, since 
\begin{equation}
    S(\mathbf{q}, t) = S^{x}(\mathbf{q}, t) + S^{y}(\mathbf{q}, t) + S^{z}(\mathbf{q}, t), 
\end{equation}
the three components $x$, $y$, $z$ of $C^{\alpha}(\mathbf{r} - \mathbf{r'}, t)$ can be calculated simultaneously within the innermost $\mathtt{for}$ loop, taking advantage of the fact that the three components of spin vectors are stored continuously in the memory. Secondly, we can compute and store $\langle S^\alpha_{\mathbf{r'}}(0)\rangle$ before the time evolution, as it remains constant throughout the simulation, eliminating the need to recalculate it at every time step.  Also, a similar strategy can be applied on $\langle S^\alpha_{\mathbf{r}}(t)\rangle$ though it needs to be recalculated for each time $t$. We perform this operation before the calculation of $S(\mathbf{q}, t)$ in the three $\mathtt{for}$ loops, and it brings obvious performance improvement. Thirdly, we optimize the memory access of $S^\alpha_{\mathbf{r'}}(0)$ by rearranging the order of data stored in main memory for $S^\alpha_{\mathbf{r'}}(0)$ to ensure that memory access in the second and third $\mathtt{for}$ loops is continuous. Specifically, we create a new array of size $3MN$ for $S^\alpha_{\mathbf{r'}}(0)$ to store the spin vectors on every site from $M$ Monte Carlo realizations continuously. For $S^\alpha_{\mathbf{r}}(t)$, we adopt a similar approach, and create a new array of size $3M$ in the first $\mathtt{for}$ loop to store spin vectors on the site $r$ continuously. Moreover, we always load data from the main memory into registers and perform calculations on registers since they are significantly faster. Our single test on 2D $100 \times 100$ square lattice with 50 MC realizations shows that these optimization strategies make the calculation of $S(\mathbf{q}, t)$ 4 times faster compared to the unoptimized implementation.


\begin{figure} 
    \centering
    \includegraphics[width=0.8\textwidth]{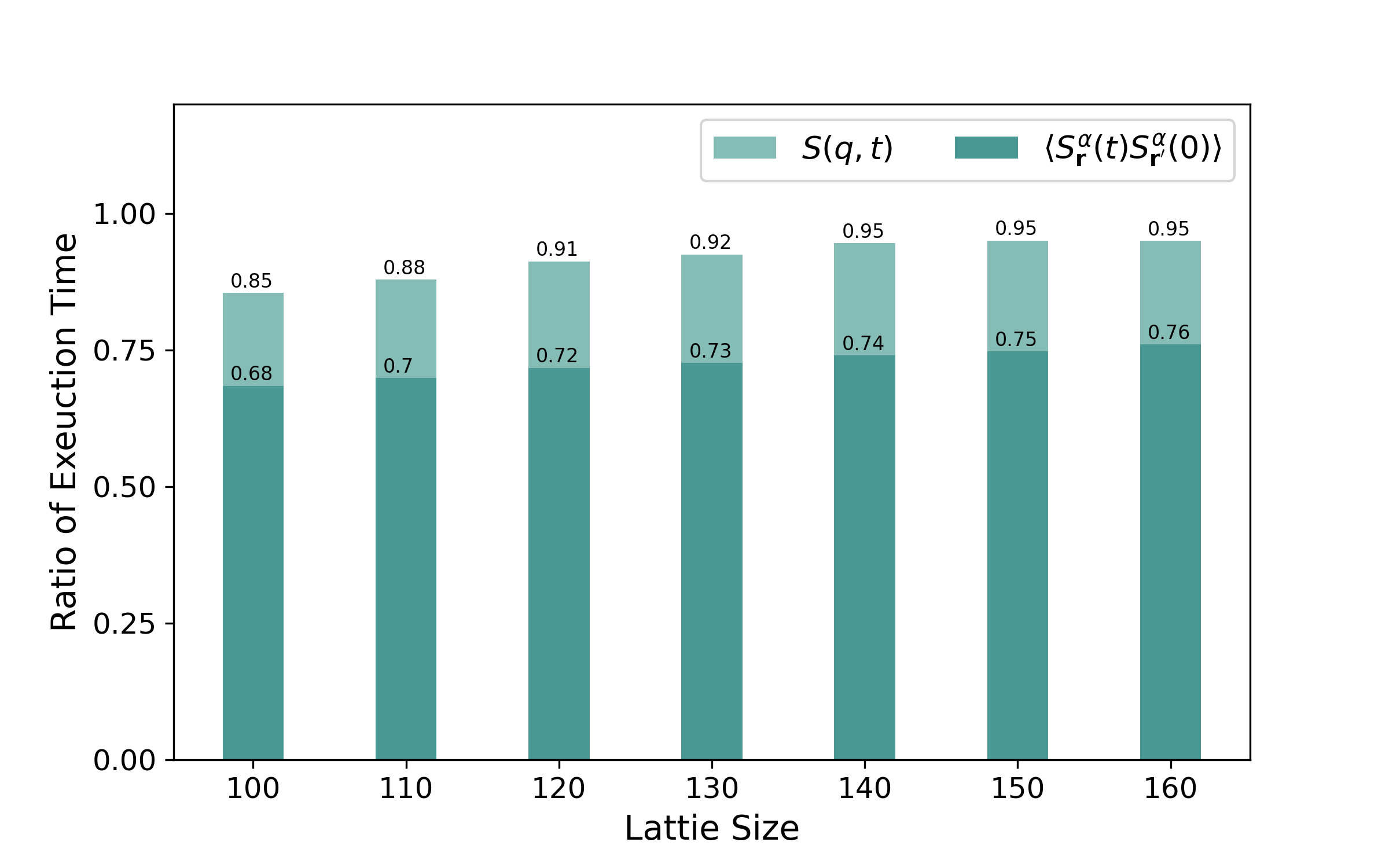}
    \caption{ We calculate the ratio between the execution time of $S(\mathbf{q}, t)$ and $\langle S^\alpha_{\mathbf{r}}(t) S^\alpha_{\mathbf{r'}}(0) \rangle$ over the total execution time on different lattice sizes L($N=L^2$). 50 MC realizations were used for the statistical average in all the implementations in this work. }
    \label{fig:ratio}
\end{figure}

\subsection{GEMM variant}

\begin{algorithm}[hbt!]
\caption{The GEMM implementation for $S(\mathbf{q}, t)$}\label{alg:baseline}
\KwData {$\mathbf{S}_{i,r}(0)$ $M$ spin configurations from Monte Carlo ($1 \leq  i \leq M$) }
\For{$r' \gets 1$ \KwTo $N$} 
{
     $\langle S^\alpha_{r'}(0)\rangle \gets \mathbf{S}_{i,r'}(0)$ ($1 \leq  i \leq M$)\Comment{Calculate statistical average}
}
Rearrange $\mathbf{S}_{i,r}(0)$  \Comment{Rearrange to use GEMM}

\For{$t \gets 1$ \KwTo $T$} 
{

\For{$i \gets 1$ \KwTo $M$}
{
    \For{$r \gets 1$ \KwTo $N$}
    {
         $\mathbf{S}_{i,r}(t) \gets \mathbf{S}_{i,r}(t-1) $ \Comment{Solve LLG equation}
    }
}

\For{$r \gets 1$ \KwTo $N$}
{
     $\langle S^\alpha_{r}(t)\rangle \gets \mathbf{S}_{i, r}(t)$ ($1 \leq  i \leq M$)  \Comment{Calculate statistical average}
}
 Rearrange $\mathbf{S}_{i,r}(t)$  \Comment{Rearrange to use GEMM}

 $C(\mathbf{r} - \mathbf{r'}, t) \gets $ GEMM($\mathbf{S}_{i,r}(t)$, $\mathbf{S}_{i,r}(0)$)

 $C(\mathbf{r} - \mathbf{r'}, t) \ -= \langle S^\alpha_{\mathbf{r}}(t)\rangle \langle S^\alpha_{\mathbf{r'}}(0)\rangle $
 
 $S(\mathbf{q}, t) \gets e^{-i\mathbf{q}\cdot (\mathbf{r} -\mathbf{r'})}  C(\mathbf{r} -\mathbf{r'}, t)$

}
\end{algorithm}

\begin{figure} 
    \centering
    \includegraphics[width=0.7\textwidth]{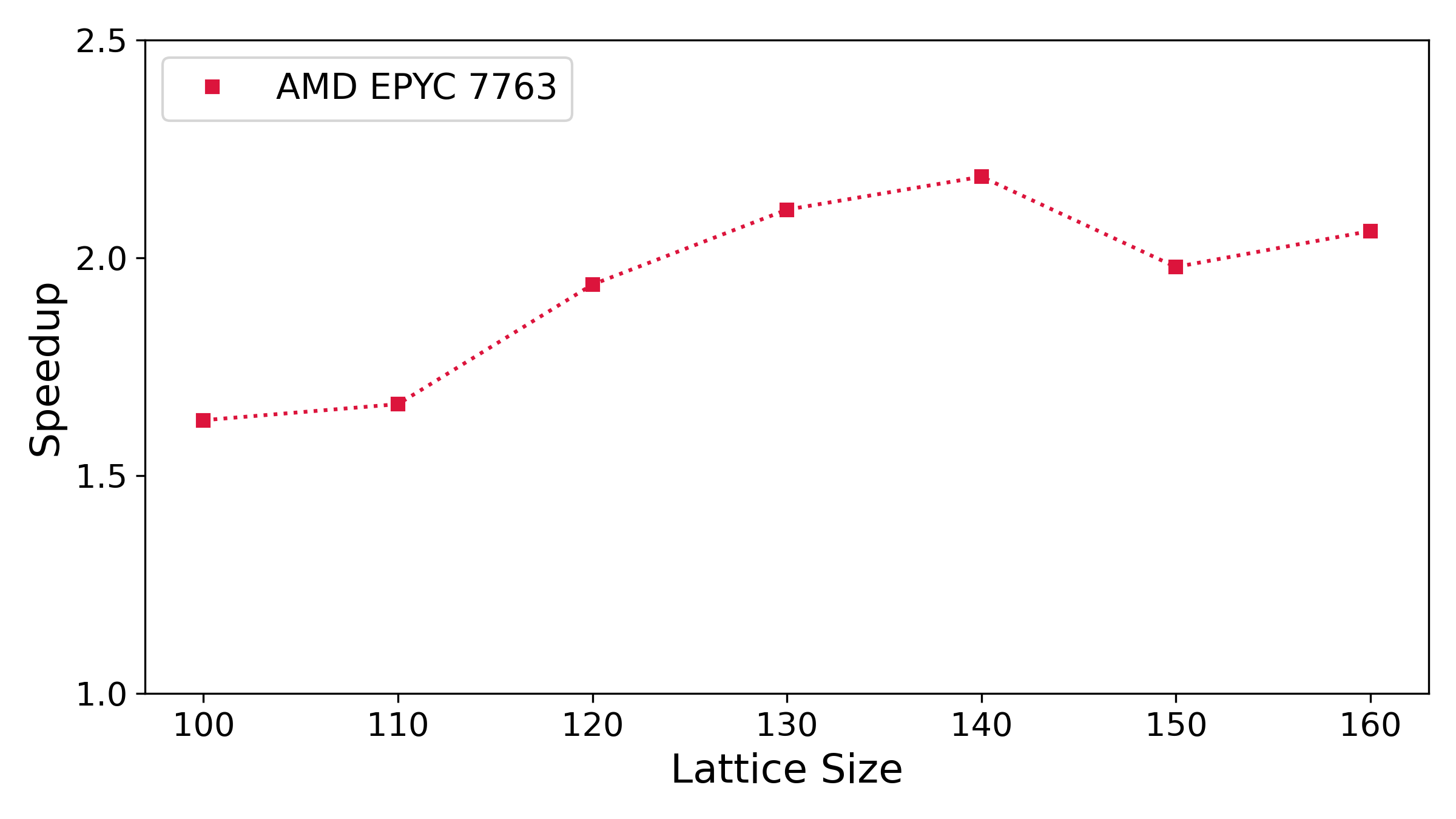}
    \caption{ The speed up of GEMM implementation over the baseline variant on different lattice sizes. OpenBLAS 0.3.20 was used in the GEMM variant.}
    \label{fig:gemm_speed_up}
\end{figure}

As shown in Fig.(\ref{fig:ratio}), even with the optimization we have applied, the computation time for calculating $S(\mathbf{q}, t)$ still dominates the overall simulation. Guided by the theoretical analysis, we also show the ratio of the execution time of $\langle S^\alpha_{\mathbf{r}}(t) S^\alpha_{\mathbf{r'}}(0) \rangle$ over the overall simulation in Fig.(\ref{fig:ratio}), finding that it takes approximately 68\% $\sim$ 76\% of the total execution time. However, we notice that the computing of $\langle S^\alpha_{\mathbf{r}}(t) S^\alpha_{\mathbf{r'}}(0) \rangle$ bears a striking resemblance to the General Matrix Multiply (GEMM) subroutine. GEMM employs a series of architecture-aware optimization strategies, such as cache- and register-level data re-use, prefetching, and vectorization that improve the hardware utilization of a program from a marginal $< 1$\% to a near-optimal efficacy ($>90\%$). To leverage the highly optimized GEMM subroutine, the order of data in memory for spin configurations $S^\alpha_{\mathbf{r}}(t)$ at time $t$ needs to be rearranged.  For each MC realization, we first store the $x$ component of spin vectors on every site, followed by the $y$ and $z$ components. The resulting rearranged spin configurations can be treated as a 2-dimensional matrix of size $N\times 3M$ stored in column-major. As for the spin configurations $S^\alpha_{\mathbf{r}}(0)$ at time zero, its data order has already been arranged in the baseline implementation, and then it can be treated as a 2-dimensional matrix of size $3M\times N$ in column-major. An additional matrix of size $N\times N$ is required to store the result of $\langle S^\alpha_{\mathbf{r}}(t) S^\alpha_{\mathbf{r'}}(0) \rangle$ computed by calling the GEMM subroutine. Another cost of using the GEMM subroutine is that the order of data for $S^\alpha_{\mathbf{r}}(t)$ needs to be rearranged for every step. The optimizations mentioned in the baseline were also integrated into the new implementation. In Fig.(\ref{fig:gemm_speed_up}), we compare the performance of the baseline with the GEMM variant on several different lattice sizes on a multi-core AMD EPYC 7763 CPU.


\subsection{Kernel fusion}

In addition to providing algorithmic designs to adapt for the highly efficient GEMM module, we also observe that the element-wise operations to compute ``Daxpy" in Eq.(\ref{eq:ssc}) and Fourier transform in Eq.(\ref{eq:sqt}) also consume non-negligible execution time, as shown in Fig.(\ref{fig:ratio}). Kernel fusion, a strategy that fuses the memory footprint of the element-wise operation with the compute-bound GEMM operation to hide the memory latency, is a sound solution that benefits a series of GEMM-based scientific computing and machine-learning applications \cite{zhai2021ft, zhai2022bytetransformer}. Therefore, we delve into the black box of GEMM kernels, enabling memory-bandwidth efficient computations for ``Daxpy" and  Fourier transforms by fusing their memory accesses with the GEMM kernel. Before detailing how we fuse the element-wise operations into the GEMM kernel, we present the rationale behind how an ordinary GEMM kernel reaches bare-metal performance.

We start our optimizations by implementing our in-house GEMM subroutine that provides performance comparable to OpenBLAS. To mitigate the memory latency for data movement between memory layers, we adopt a series of delicate optimizations including cache blocking, packing, vectorization, and prefetching at the assembly code level. The outermost three layers of the $\mathtt{for}$ loop are partitioned to allow submatrices of $A$ and $B$ to reside in specific cache layers. The step sizes of these three $ \mathtt{for}$ loops, $\mathtt{M_C}$, $\mathtt{N_C}$, and $\mathtt{K_C}$, define the size and shape of the macro kernel, which are determined by the size of each layer of the cache. Loading data from the main memory needs to access the translation lookaside buffer (TLB) to locate the page table entry of the data chunk. In practice, a miss in the TLB level leads to twice main memory accesses and accordingly a latency of hundreds of nanoseconds, which can significantly slow down a high-performance compute kernel. Therefore, we pack the data blocks of submatrices $A$ and $B$ into contiguous memory buffers $\tilde{A}$ and $\tilde{B}$ such that the cost at the TLB level is minimized. Once the memory buffers are packed, a macro kernel updates an $M_C\times N_C$ submatrix of C by iterating over $A$ $(M_R\times K_C)$ multiplying $B$ $(K_C\times N_R)$ in the assembly microkernel, where prefetch instructions are carefully scheduled by considering the hardware and software parameters. Considering the common hardware specifications of AVX2-enabled processors, we choose a static parameter setup for $\{M_C, N_C, K_C\}$ =$\{512, 9216, 240\}$ and $\{M_R, N_R\}$ =$\{4, 12\}$.

Rather than directly storing the $(M_R\times K_R)$ $C$ submatrix into the main memory, we conduct fused ``Daxpy" and Fourier transform computations when these data remain in registers. In practice, our microkernel computes a $(4\times 12)$ $C$ submatrix in an outer-product manner. These submatrix elements are held in twelve single-instruction multiple-data (SIMD) registers until the matrix multiplication algorithm loops over in the shared dimension $k$. In the fused implementation, we pass two additional pointers into the assembly microkernel, such that the $(4\times 12)$ submatrix elements can be re-used at the register level. From a global perspective, we have reduced costs by at least a $N^2$ times in the main memory by enabling the kernel fusion strategy. From the perspective of the assembly level, the kernel fusion strategy has exhausted the hardware resources for both integer and SIMD registers. Since an AVX2-enabled processor consists of 16 256-bit $\mathtt{ymm}$ SIMD registers, an ordinary $4\times12$ AVX2 kernel consumes 15 SIMD registers: 12 SIMD registers to hold $C$ elements, 1 register to hold an $A$ tile and 2 registers to hold broadcast $B$ tiles. As the outer-product-based micro kernel loops over the $k$ dimension, the three SIMD registers allocated to hold $A$ and $B$ tiles are liberated. In our self-implemented fused kernel, 2 of the free SIMD registers are allocated to hold intermediate shuffle results and the remaining 2 free SIMD registers are utilized to hold data tiles of the fused matrices. It is also worth mentioning that the memory accesses to fused matrices are also incorporated with delicate assembly-level prefetching to mitigate the memory latency. Therefore, the fused operation purely introduces computational overhead, which is negligible to the baseline cubical GEMM operation.

\begin{figure} 
    \centering
    \includegraphics[width=0.7\textwidth]{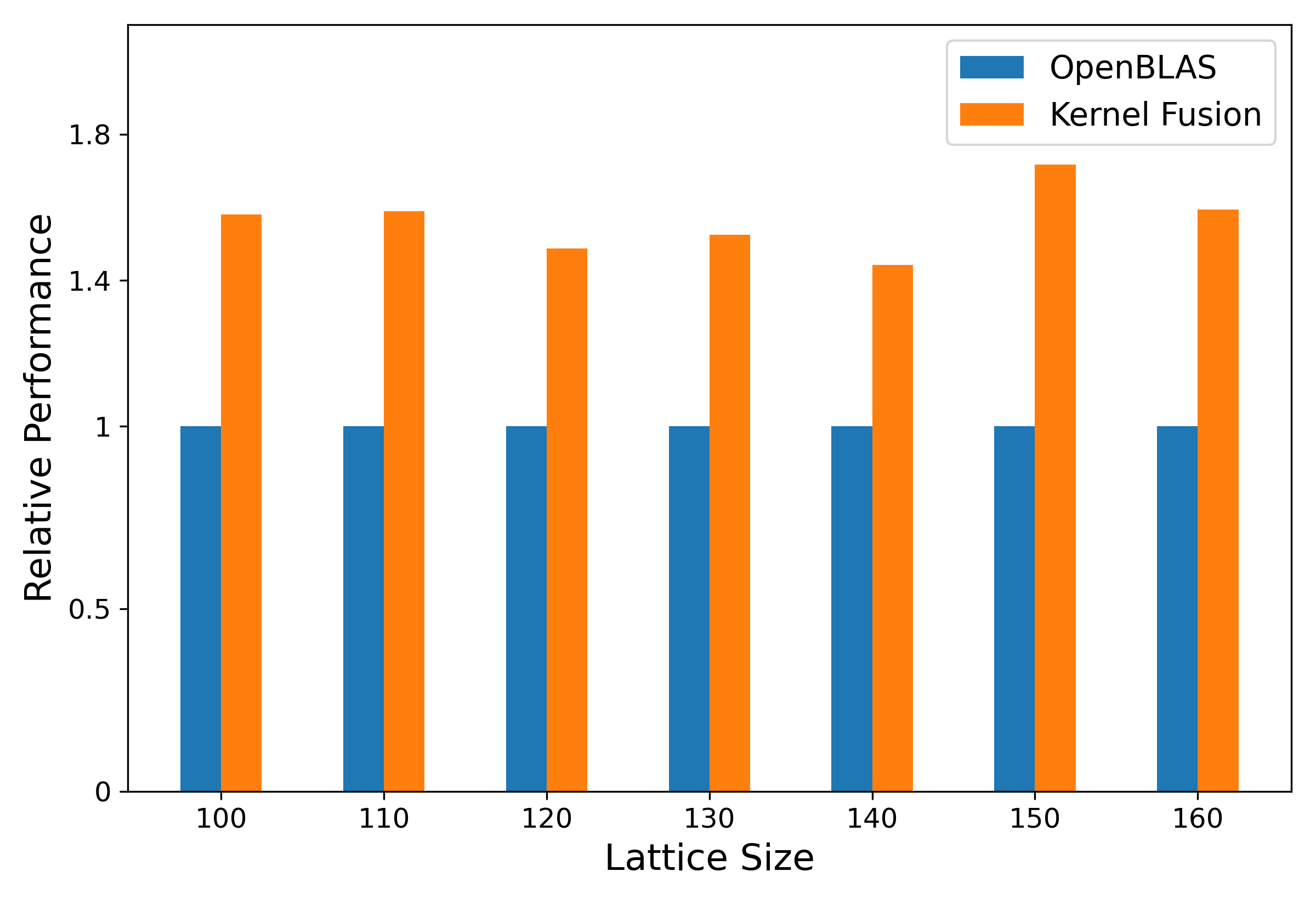}
    \caption{ The relative performance improvement of kernel fusion over the GEMM variant on different lattice sizes L($N=L^2$). }
    \label{fig:fusion_perf}
\end{figure}

\subsection{Parallel Design}

In addition to introducing data-level and instruction-level parallelism for the single-thread GEMM implementation with kernel fusion, we further extend our algorithm to embrace the architecture of high-performance multi-node computer clusters. To be more specific, we target the Non-Uniform Memory Access (NUMA) computer clusters, where each core contains its own memory controller and can communicate with other nodes through a high-performance bus. Since the compute cores of a NUMA cluster share a huge L3 cache but possess their private L1 cache, a parallel implementation must honor this hardware feature to ensure the cache locality and high performance in the GEMM kernel. In fact, our parallel GEMM kernel partitions the workload from the $N$ dimension such that the $\tilde{B}$, which resides in the shared L3 cache, is naturally parallelized without interfering with the NUMA cache architecture. Regarding the $\tilde{A}$ data chunk, which resides in the distinct L1 cache, we present a parallel copy scheme, where each thread copies one's own $A$ slice into its nearest cache chunk. This parallel design not only maximizes the hardware utilization regarding parallelism but maintains the cache policy of NUMA architecture and accordingly the scalability on a multi-core NUMA cluster as well. 

Built upon the ordinary GEMM implementation, the parallel algorithm leads to minimal updates on the fused assembly kernel. Compared with the single-thread fused variant, a parallel implementation reduces the $4\times12$ matrix to a distinct data address of each thread. Once the parallel fused GEMM kernel is completed, a further round of reduction among all the threads is conducted, such that the final result is obtained. The cost of the above final-round reduction is linear to the number of compute cores so is negligible to the end-to-end performance.

\begin{figure} 
    \centering
    \includegraphics[width=\textwidth]{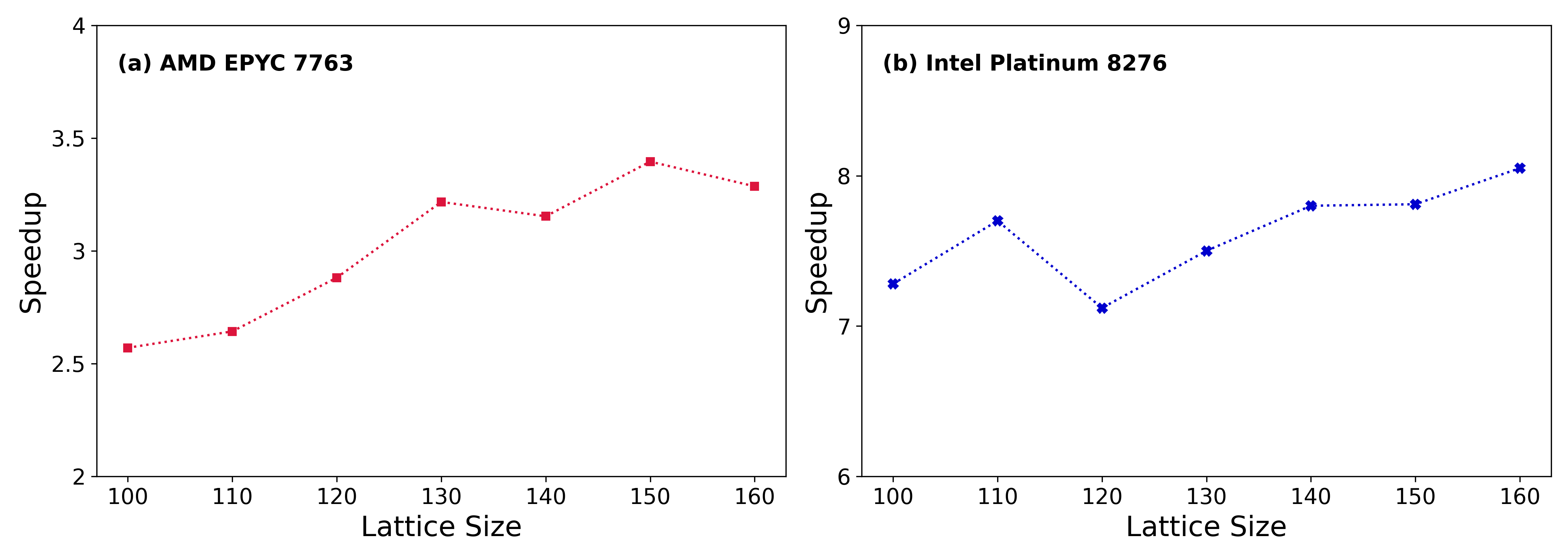}
    \caption{ The speedup of kernel fusion compared to the baseline variant on varying lattice sizes L ($N=L^2$) on both AMD and Intel platforms.}
    \label{fig:speed_up}
\end{figure}


\section{Results}

\begin{figure} 
    \centering
    \includegraphics[width=\textwidth]{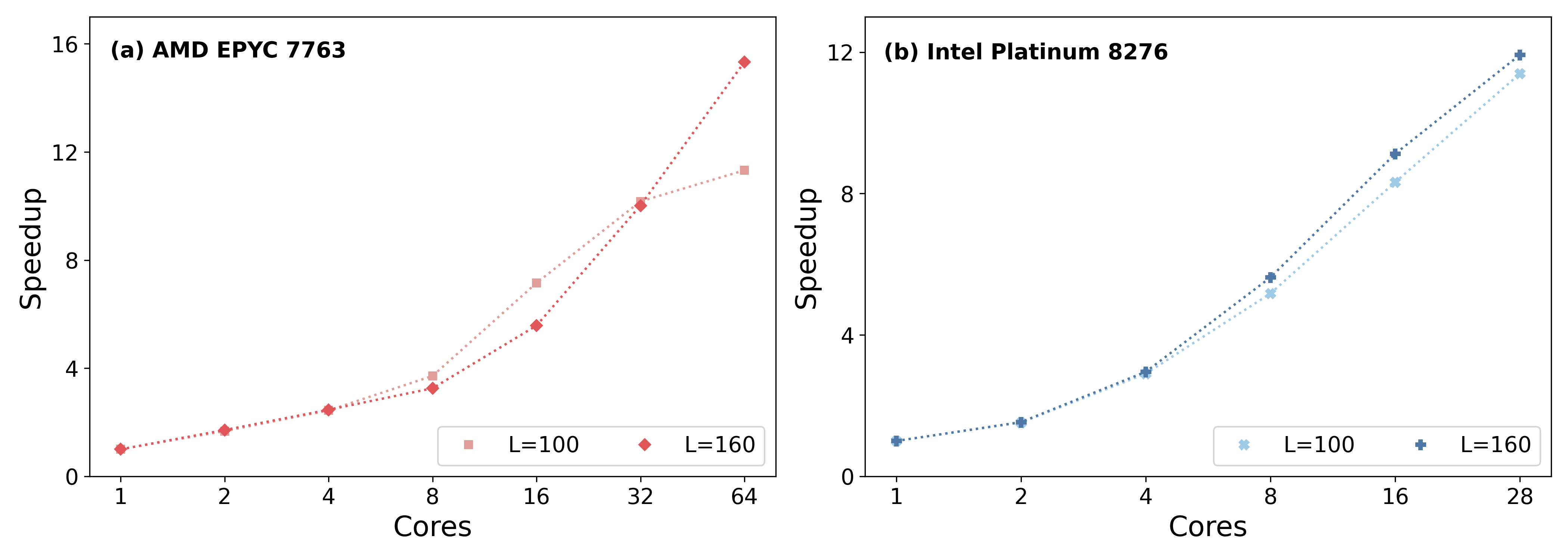}
    \caption{ The scalability of our ``kernel fusion" implementation on both AMD and Intel platforms. The speedup was calculated by comparing the execution time on multi-CPU cores and a single CPU core.  }
    \label{fig:speed_core}
\end{figure}

Fig.(\ref{fig:fusion_perf}) shows the performance improvement of kernel fusion on varying lattice sizes using  AMD EPYC 7763 CPU. We observe that kernel fusion can increase the performance by $44\% \sim 71\%$ compared to the GEMM variant. In Fig.(\ref{fig:speed_up}), we calculate the speedup of the ``kernel fusion'' implementation over the baseline variant on both AMD and Intel platforms. A 28-core Intel Platinum 8276 processor, whose boost frequency is up to 4.0 GHz, was used for evaluation on the Intel platform. It is worth mentioning that our GEMM kernel is currently yet to support the AVX-512 instruction set, which is only available on Intel CPUs now. On both the Intel and AMD processors, the fully optimized variant achieves significant speedup over the threaded baseline for the atomistic spin dynamics simulation.  Fig.(\ref{fig:speed_core}) presents the speed-up of our optimal variant on multi-core processors compared to single-core, which demonstrates the sound scalability of our fused GEMM.

\begin{figure} 
    \includegraphics[width=\textwidth]{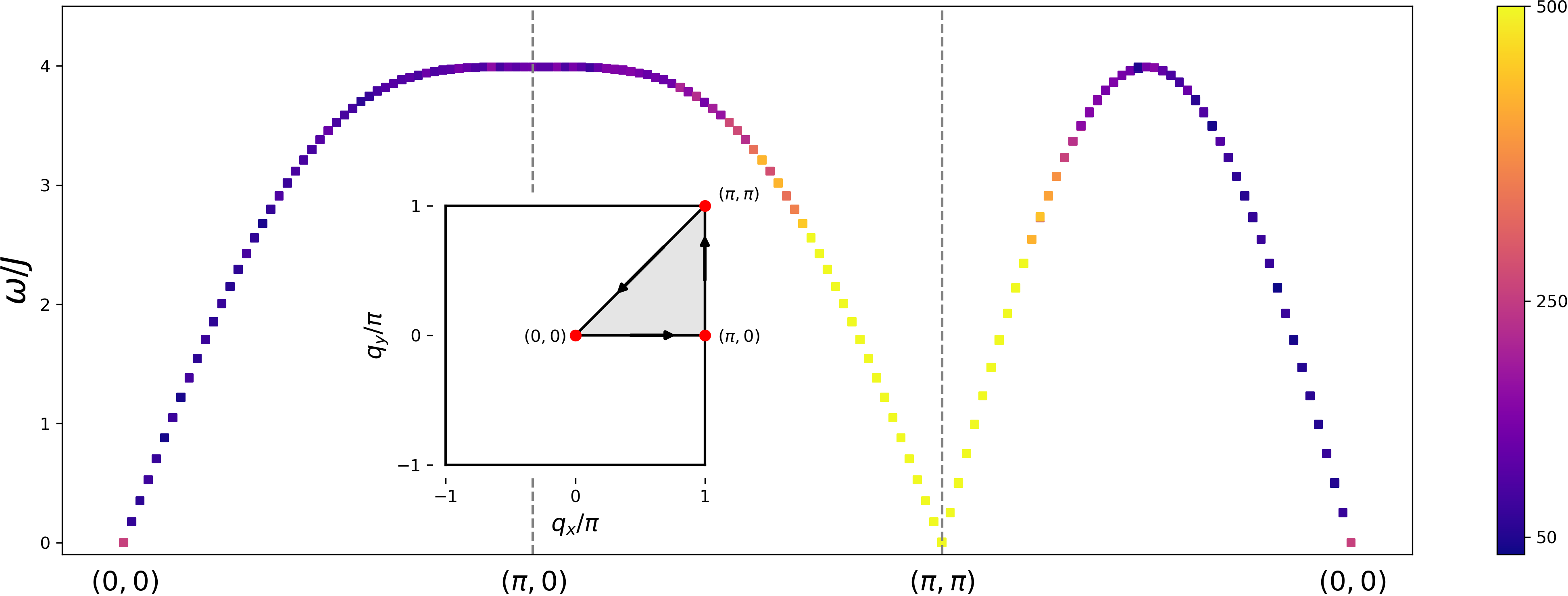}
    \caption{ The dynamical spin structure factor $S(q, \omega)$ calculated in the anti-ferromagnetic phase at temperature $T=0.01J$, along the momentum path $(0, 0) \rightarrow (\pi, 0) \rightarrow (\pi, \pi) \rightarrow (0, 0)$. The subfigure shows the high symmetry path of the first Brillouin zone for the square lattice.}
    \label{fig:sqw}
\end{figure}

Moreover, we apply our implementation of spin dynamics simulation to calculate the dynamical spin structure factor $S(q, \omega)$ for the anti-ferromagnetic Heisenberg model. Using Monte Carlo and simulated annealing, we obtain 50 spin configurations of the 2-dimensional square lattice of size $100\times 100$ with periodic boundary conditions. The simulated annealing starts from a high temperature $T_0=30J$, and then gradually lowers it to the target temperature $0.01J$ using $T_{i+1} = 0.99T_i$. At each temperature, $10^5$ Monte Carlo sweeps ($10^9$ spin flips) are performed, and additional $5\times 10^5$ Monte Carlo sweeps were used at the target temperature to reach equilibrium.  The dynamical spin structure factor $S(\mathbf{q}, \omega)$ was calculated along the high symmetry path of the first Brillouin zone using Eq.(\ref{eq:ll}) and Eq.(\ref{eq:sqt}), as shown in Fig.(\ref{fig:sqw}). 

\section{Conclusion}

In this work, we explore the high-performance implementations of atomistic spin dynamics simulations. Specifically, we find that the calculation of the spin structure factor $S(\mathbf{q}, t)$ consumes the vast majority of the simulation time. We identify  the term $\langle S^\alpha_{\mathbf{r}}(t) S^\alpha_{\mathbf{r'}}(0) \rangle$ as the most computationally expensive part. To optimize the overall performance, we propose to use the GEMM subroutine to calculate $\langle S^\alpha_{\mathbf{r}}(t) S^\alpha_{\mathbf{r'}}(0) \rangle$ by rearranging the spin configurations in the memory during time evolution.  The highly optimized GEMM subroutine can achieve near-optimal hardware utilization, and speed up the simulation significantly, especially on computing nodes with a small number of CPU cores. Furthermore, we fuse the element-wise operation in the calculation of $S(\mathbf{q}, t)$ into our in-house GEMM kernel to hide the memory latency of transferring the covariance matrix. On the Intel platform, we achieve a speedup of up to 8-fold compared with the baseline variant.  Also, we highlight the fact that the optimizations we propose can be extended to simulate the dynamical properties of any classical spin Hamiltonian, regardless of the geometry of the lattice.


\section*{Acknowledgments}

This work is supported by the U.S. Department of Energy, Office of Science, Basic Energy Sciences under Award No. DE-SC0022216. Part of this work was also supported under Contract DE-AC02-76SF00515, both for the Materials Sciences and Engineering Division and for the Linac Coherent Light Source (LCLS). The numerical calculations were performed using the resources of the National Energy Research Scientific Computing Center (NERSC), a DOE Office of Science User Facility operated under Contract No. DE-AC02-05CH11231.

\appendix

 \bibliographystyle{elsarticle-num} 





\end{document}